\documentclass[12pt]{article}
\begin{document}

\begin{center}
{\large \bf Hypothesis of a new fundamental interaction versus the particle oscillation concept}
\vspace{0.5 cm}

\begin{small}
\renewcommand{\thefootnote}{*}
L.M.Slad\footnote{slad@theory.sinp.msu.ru} \\
{\it Skobeltsyn Institute of Nuclear Physics,
Lomonosov Moscow State University, Moscow 119991, Russia}
\end{small}
\end{center}

\vspace{0.3 cm}

\begin{footnotesize}
\noindent                    
{\bf Abstract.} A significant part of the present work is devoted to proving that the basic elements of the particle oscillation concept either do not correspond to the principles of classical logic, or violate the energy-momentum conservation law, or contradict the quantum mechanical basis of coherence, or represent a primitive falsehood. Our analysis concerns successively all stages of the formation of the concept from its conceiving to the assertion about the conversion of the solar electron neutrino into a muon one. When discussing a new fundamental interaction, we note the decisive role in the outcome of the processes of changing the handedness of a neutrino (antineutrino) at each act of its interaction with a real or virtual massless pseudoscalar boson, due to which, at the exit from the Sun, the fluxes of left- and right-handed electron neutrinos become approximately equal. Thanks to the new interaction, beta decays of nuclei have a mode with the emission of a massless pseudoscalar boson, at which the antineutrino changes its handedness and becomes unobservable, what is the essence of the reactor antineutrino anomaly. 
\end{footnotesize}

\vspace{0.5 cm}

\begin{small}

\begin{center}
{\large \bf 1. Introduction}
\end{center}

The emergence of the solar neutrino problem is caused by the report \cite{1} about the absence of observed transitions of $^{37}$Cl into $^{37}$Ar under the action of such neutrinos. It have been stated the upper limit for such a transition rate as 3 SNU (1 SNU is 10$^{-36}$ captures per target atom per second), while the rate predicted by Bahcall \cite{2} was 30$^{+30}_{-15}$ SNU.

The reaction to this report was well conveyed by Reines \cite{3}: "It is interesting to note that if a positive result were obtained in the Davis experiment, we would suddenly face the problem of whether it is due to the Sun or something else. A negative result is important because that it allows us to assert, in the event of a further positive result in a neutrino experiment, that the effect is caused precisely by solar neutrinos.This is significant, because the Universe is full of surprises ... When in February 1972 we discussed this issue at a conference in Irvine, it was interesting to observe how, in search of the reasons for the discrepancy, astrophysicists pointed to specialists in the nucleus, the latter to neutrino physicists, and those in turn to astrophysicists". These words were said by Reines during the discussion of Pontecorvo's report at the seminar on the $\mu - e$ problem. It is noteworthy that Reines does not say a word about the solar neutrinos oscillations, the assumption of which was put forward three years earlier \cite{4}. This indicates that, for the time being, the neutrino oscillation concept was weakly perceived by the scientific community.

Over time, Wolfenstein \cite{5} extended it to the transformation of a neutrino as it moves in a homogeneous medium. Seven years later, Mikheev and Smirnov \cite{6} announced the extension of the Wolfenstein equation to the inhomogeneous medium of the Sun, but did not describe the evolution of the state of the solar neutrino as it moves from its production place to an installation on Earth. Such a description with the oscillation parameters given by the Super-Kamiokande and SNO collaborations was partly made in \cite{7} by analytical and numerical analysis of this equation. The main task of \cite{6} was to argue the statement about the transformation of an electron neutrino into a muon neutrino during its passage through the Sun. The proposed arguments, as we will prove below, are false. Despite the initial perception of the work by reviewers as erroneous, over time, the belief that the Mikheev-Smirnov-Wolfenstein mechanism is able to solve the solar neutrino problem has become a fairly widespread phenomenon among the scientific community.

Having analyzed all stages of the particle oscillation concept from its inception, which concerned neutral $K$-mesons, to the appearance of the Mikheev-Smirnov-Wolfenstein mechanism, we find either ignoring the rule of sufficient reason, or ignoring the law of conservation of energy-momentum, or neglecting the quantum mechanical basis of coherence , or elementary falsity. This will inevitably lead sooner or later to the collapse of the concept of particle oscillations. I hope that it will be helped, in particular, by the emergence of a logically simple solution to the solar neutrinos problem. The main theseses of such a solution and their correspondence with experiments were outlined in February 2015 in my work \cite{8}.

It would seem that the whole history of physics pointed to the fact that some new for us, rather hidden interaction is responsible for the emergence of the solar neutrino problem. Surprisingly, that, as it is clear from the utterance of Reines \cite{3}, the question of hidden interaction was not even raised at the Irvine conference. My hypothesis in work \cite{8} postulates the existence of an interaction whose carrier is a massless pseudoscalar boson that has a Yukawa coupling with at least the electron neutrino, proton, and neutron, but not with electrons. Due to this interaction, neutrinos during  their movement inside the Sun experience about a dozen collisions with nucleons. At each such collision, the neutrino handedness changes from left to right and vice versa, so that at the exit from the Sun the ratio of left- and right-handed electron neutrino fluxes is 0.515:0.485. In addition, at each collision of a neutrino with nucleons, its energy decreases slightly. Our model with one free parameter provides good agreement between theoretical and experimental results for the rates of all five observed processes with solar neutrinos \cite{9}, \cite{10}.

This paper is devoted, on the one hand, to the analysis of the particle oscillation concept from its inception and, on the other hand, to a discussion of separate elements of the postulated new interaction.

\begin{center}
{\large \bf 2. The particle oscillation concept}
\end{center}
\begin{center}
{\bf 2.1. Neutral K-mesons}
\end{center}

The discovery of hyperons and $K$-mesons was an unexpected and amazing surprise, as it is said in the monograph of that time \cite{11}.

In the emerging theory of new unstable particles, first of all, ones tried to transfer such notions of symmetry as isospin and charge conjugation, attributed to stable particles. Gell-Mann \cite{12} supposed that the new unstable particles are fermions with integer isospin and bosons with half-integer isospin and have antiparticles that differ from them. Gell-Mann and Pais [13] believed that strong interactions are responsible for all productions of hyperons and $K$-particles, while weak interactions are responsible for their decays. Therefore, in their assertion, the law of conservation of isospin allows the reaction of production of $\theta^{0}$-meson with the projection of isospin -1/2 (hereinafter referred to as $K^{0}$)
\begin{equation}
\pi^{-} + p \rightarrow \Lambda^{0} + \theta^{0},
\label{1}
\end{equation}
but prohibits the process of production of its antiparticle
\begin{equation}
\pi^{-} + p \rightarrow \Lambda^{0} + \bar{\theta}^{0}.
\label{2}
\end{equation}
At the same time, due to the violation of the law of conservation of isospin in weak interactions, decays of the $\theta^{0}$-particle and $\bar{\theta}^{0}$-antiparticle are admissible according to the same schemes:
\begin{equation}
\theta^{0} \rightarrow \pi^{+} + \pi^{-}, \qquad \bar{\theta}^{0} \rightarrow \pi^{+} + \pi^{-}.
\label{3}
\end{equation}

The relations (\ref{3}) allow the virtual transition $\theta^{0} \leftrightarrow \pi^{+}+\pi^{-} \leftrightarrow \bar{\theta}^{0}$, i.e. the particle $\theta^{0}$ and its antiparticle $\bar{\theta}^{0}$ do not completely differ from each other. Gell-Mann and Pais find it convenient to introduce two quanta $\theta^{0}_{1}$ and $\theta^{0}_{2}$ (hereinafter referred to as $K^{0}_{1}$ and $K^{0}_{2}$, and then as $K^{0}_{S}$ and $K^{0}_{L}$), transforming into themselves upon charge conjugation:
\begin{equation}
\theta^{0}_{1} = (\theta^{0} + \bar{\theta}^{0})/\sqrt{2}, \qquad \theta^{0}_{2} = (\theta^{0} - \bar{\theta}^{0})/i\sqrt{2}.
\label{4}
\end{equation}
Since each of the quanta $\theta^{0}_{1}$ and $\theta^{0}_{2}$ can be assigned its own lifetime and mass, Gell-Mann and Pais consider them to be true "particles", and at the same time they treat $\theta^{0}$ and $\bar{\theta}^{0}$ as a "mixture of particles".

It is incorrect to interpret relations (\ref{4}) as an expression of the wave function of one of the particles in terms of the superposition of the wave functions of two other particles, which takes place in \cite{14}. Since the values of the wave functions are $c$-numbers, it is easy to find their sum (difference), but in this sum they lose their individuality and it is impossible to separate them from the sum and perform their separate transformations.

I find it useful to express here the essence of the addition of "particles" or "mixture of particles" in terms of known logical operations. Let us introduce the notion of a linear space of particles over the field of functions of space-time coordinates, $\cal{L}$. Each particle is associated with its own basis vector in the space $\cal{L}$, denoted by the name of the particle. Any particle state and its time transformations are displayed in $\cal{L}$ by vectors obtained by multiplying its basis vector by an appropriate wave function. Each mixture of particles is associated with only one vector in the space $\cal{L}$, which is given by a well-defined decomposition into the basis vectors of this space at the initial moment and is denoted by the name of the mixture. The time transformation of each mixture is given by the transformation of the states of the particles, into whose basis vectors it is decomposed. Finally, the state vectors of the particles at time $t$ are decomposed into the initial mixture vectors. The coefficients of these decompositions play the role of the probability amplitudes for the existence of the corresponding mixtures at time $t$.

The time transformation of the wave function of a particle with mass $m$ is well known: its initial value acquires an additional phase factor $\exp(-iEt)$, where $E=\sqrt{m^{2}+ p^{2}}$, $p$ is momentum modulus. The question of the time dependence of the wave function of a "mixture of particles" was considered, in the spirit of what was said above, in the work of Pais and Piccioni \cite{14}. From the relations (\ref{4}) ones obtain
\begin{equation}
\theta^{0} = (\theta^{0}_{1} + i\theta^{0}_{2})/\sqrt{2}, \qquad \bar{\theta}^{0} = (\theta^{0}_{1} - i\theta^{0}_{2})/\sqrt{2}.
\label{5}
\end{equation}
Getting the time transformation $\theta^{0}$ is like this:
\begin{equation}
\theta^{0} \rightarrow (\theta^{0}_{1}\exp(-iE_{1}t) + i\theta^{0}_{2}\exp(-iE_{2}t))/\sqrt{2} = 
((\theta^{0} + \bar{\theta}^{0})\exp(-iE_{1}t) + (\theta^{0} - \bar{\theta}^{0})\exp(-iE_{2}t))/2.
\label{6}
\end{equation}

Hence, under the condition $p_{1}=p_{2}\equiv p$, $E_{1} \gg m_{1}$ and $E_{2} \gg m_{2}$, the probability of existence of a meson $\theta^{0}$ at time $t$ is an oscillating quantity
\begin{equation}
P(\theta^{0}, t) \approx \cos^{2}\frac{(m_{1}^{2}-m_{2}^{2})}{4p}t.
\label{7}
\end{equation}

In \cite{13} and \cite{14}, not a single word is said about the feasibility of the law of conservation of energy-momentum in the process (\ref{1}), provided that the $\theta^{0}$-meson is a mixture of two "true" particles, $\theta^{0}_{1}$ and $\theta^{0}_{2}$ ($K_{S}$ and $K_{L}$, in current notation) with different masses $m_{1}$ and $m_{2}$. Our statement is that the feasibility of the law of conservation of energy depend on the ratio between the total uncertainty in masses (the sum of decay widths) of unstable particles $\sum_{i} \Gamma_{i}$ participating in the process (\ref{1}) and the mass difference $|m_{1} - m_{2}|$. If the total mass uncertainty exceeds this mass difference, then the feasibility of the law of conservation of energy in the process (\ref{1}) with the $\theta^{0}$-meson as a mixture of two particles is admissible. Otherwise, the produced neutral $K$-meson is forbidden to be a mixture of particles. It must itself be a particle with a dominant decay channel (\ref{3}).

Let us now turn to the numbers characterizing the process (\ref{1}). We take into account that the particle decay width $\Gamma$ and its lifetime $\tau$ are related by the equality $\Gamma \tau = 1$. The lifetimes of particles interesting to us are taken from review \cite{15}. We have: $\tau_{\pi^{-}} = 2.60 \cdot 10^{-8}$ s, $\Gamma_{\pi^{-}} = 2.53 \cdot 10^{-8}$ eV; $\tau_{\Lambda} = 2.63 \cdot 10^{-10}$ s, $\Gamma_{\Lambda} = 2.50 \cdot 10^{-6}$ eV.  So, the total mass uncertainty $\sum_{i} \Gamma_{i}$ in the process (\ref{1}) is $2.55 \cdot 10^{-6}$ eV. At the same time, the difference between the masses $m_{K_{L}} - m_{K_{S}}$ of two neutral $K$ mesons as "true" particles, the mixture of which forms $\theta^{0}$-meson, is less than this value. It is \cite{15} $3.48 \cdot 10^{-6}$ eV.

Thus, the requirement of feasibility of the law of conservation of energy does not alloow the two neutral $K$-mesons, $K^{0}$ and $\bar{K}^{0}$, to be mixtures of the two other $K$-mesons , $K^{0}_{S}$ and $K^{0}_{L}$.

Let us now recall that the appearance of the assertion about the oscillation of neutral $K$ mesons was caused by the assertion of Gell-Mann-Pais that, due to the conservation of isospin, the state of the $\theta^{0}$-meson, produced in the process (\ref{1}), has an isospin projection of -1/2. But, from the relation (\ref{6}), it follows that after an arbitrarily small interval of time after production this state does not have a definite value of the isospin projection. It seems natural to recognize the assertion about the conservation of isospin in the process (\ref{1}) as untenable and to accept that one of the admissible $K$-mesons, produced in the process (\ref{1}) and decaying into $\pi^{+} + \pi^{-}$, is a $\theta^{0}_{1}$($K_{S}$)-meson coinciding with its antiparticle. Some reasoning of Gell-Mann-Pais give arguments in favor of the admissibility of the production in the process (\ref{1}) of a $\theta^{0}_{2}$($K_{L}$)-meson, which decays into three pi-mesons.

From the four neutral $K$-mesons considered in \cite{12}-\cite{14}, only two mesons should be left in the final theory, namely, $K_{S}$ and $K_{L}$. Their Yukawa interaction with baryons should be represent as
\begin{equation}
L = g_{1}\bar{n}\gamma^{5}\Lambda K_{S}+g_{2}\bar{n}\gamma^{5}\Lambda K_{L}+
g_{3}\bar{n}\gamma^{5}\Sigma^{0} K_{S}+g_{4}\bar{n}\gamma^{5}\Sigma^{0} K_{L}+
g_{5}\bar{p}\gamma^{5}\Sigma^{+} K_{S}+g_{6}\bar{p}\gamma^{5}\Sigma^{+} K_{L},
\label{8}
\end{equation}
where pairs of constants $g_{1}$, $g_{2}$; $g_{3}$, $g_{4}$ and $g_{5}$, $g_{6}$ have close but different values, reflecting $CP$-symmetry breaking. It follows from here that at each collision of the meson $K_{L}$ ($K_{S}$) with the nucleons of matter, either a meson $K_{L}$ ($K_{S}$) is born again, or a meson $K_{S}$ ($K_{L}$) is born and then one speaks about the regeneration of $K_{S}$ ($K_{L}$) from $K_{L}$ ($K_{S}$)-meson. This regeneration refers to a number of such standard processes as the Compton effect and $\pi$-meson recharging.

The situation with two, not more, neutral $K$-mesons with certain masses would correspond to the assertions of great thinkers about sufficient reason and, in particular, to the assertion of Isaac Newton: {\it We are to admit no more causes of natural things than such as are both true and sufficient to explain their appearances. To this purpose the philosophers say that Nature does nothing in vain, and more is in vain when less will serve; for Nature is pleased with simplicity, and affects not the pomp of superfluous causes} \cite{16} (Book III. Rule I.)

\begin{center}
{\bf 2.2. Gribov-Pontecorvo concept of neutrino oscillations}
\end{center}

Soon after the publication of the results of Davis et al. \cite{1}, Gribov and Pontecorvo paper \cite{4} appeared, in which the lower value of the solar electron neutrino flux near the Earth's surface compared to that expected on the basis of the standard solar model is explained by solar neutrino oscillations, due to which some of them at the location of the experimental setup emerge as muon neutrinos.

Gribov and Pontecorvo, in their substantiation of the existence of neutrino oscillations, tend to adhere to the Gell-Mann-Pais-Piccioni procedure for neutral $K$-mesons. But if in Gell-Mann, Pais and Piccioni the elements of the procedure have clear experimental and worldview sources, then in Gribov and Pontecorvo they are taken "from the ceiling", from nowhere. Thus, in order to supplement the family of known neutrinos $\nu_{e}$ and $\nu_{\mu}$ with two new neutrinos, Gribov and Pontecorvo, adding off-diagonal terms to the mass part of the neutrino Lagrangian, symbolizing the transitions in vacuum of an electron neutrino to a muon neutrino and vice versa , give the Lagrangian the following form
\begin{equation}
L_{\rm int} = a_{1} \bar{\nu}_{e} \nu_{e} + a_{2} \bar{\nu}_{\mu} \nu_{\mu} + a_{3} \bar{\nu}_{e} \nu_{\mu} + {\rm H.c.},
\label{9}
\end{equation}
where $a_{1}$, $a_{2}$ and $a_{3}$ are constants. Diagonalization of the Lagrangian (\ref{9}) leads to the appearance of two states
\begin{equation}
\nu_{1} = \cos \theta \cdot \nu_{e} - \sin \theta \cdot \nu_{\mu}, \qquad
\nu_{2} = \sin \theta \cdot \nu_{e} + \cos \theta \cdot \nu_{\mu},
\label{10}
\end{equation}
whose masses and the value of the mixing angle are given by the constants $a_{i}$, $i=1,2,3$.

We note here that earlier, when Weinberg constructed the electron gauge theory [17], the off-diagonal mass term formed by the gauge fields already appeared in the Lagrangian. As a result of the diagonalization of his Lagrangian, Weinberg obtained the photon and massive boson fields, which he assigned the status of quantized fields. He then expressed through them the initial gauge fields, which he got rid of in the final theory. The number of initial and final entities in Weinberg remained the same. As a result, we have a phenomenal theory of electroweak interactions.

While Weinberg naturally avoided the doubling of entities, Gribov and Pontecorvo deliberately introduced such doubling, which became the cornerstone of their concept of neutrino oscillations.
The states $\nu_{1}$ and $\nu_{2}$ have been given status, using Gell-Mann-Pice terminology, true particles, neutrinos $\nu_{e}$ and $\nu_{\mu}$ have been given mixture status: 
\begin{equation}
\nu_{e} = \cos \theta \cdot \nu_{1} + \sin \theta \cdot \nu_{2}, \qquad
\nu_{\mu} = -\sin \theta \cdot \nu_{1} + \cos \theta \cdot \nu_{2}.
\label{11}
\end{equation}

Applying the Pais-Piccion procedure to the relations (\ref{11}) and (\ref{10}) regarded to the initial moment of time, we have
$$
\nu_{e} \rightarrow \cos \theta \cdot \nu_{1}\exp(-iE_{1}t) + \sin \theta \cdot \nu_{2}
\exp(-iE_{2}t)= $$
\begin{equation}
\cos \theta (\cos \theta \cdot \nu_{e} - \sin \theta \cdot \nu_{\mu})\exp(-iE_{1}t) + \sin \theta (\sin \theta \cdot \nu_{e} + \cos \theta \cdot \nu_{\mu})\exp(-iE_{2}t).
\label{12}
\end{equation}

Hence, under the condition $p_{1}=p_{2}\equiv p$, $E_{1} \gg m_{1}$ and $E_{2} \gg m_{2}$, the probability of detecting neutrinos $\ nu_{e}$ at time $t$ is an oscillating quantity:
\begin{equation}
P(\nu_{e}, t) = 1 - \sin^{2} 2\theta \sin^{2}\frac{(m_{1}^{2}-m_{2}^{2})}{4p}t.
\label{13}
\end{equation}
The length of its oscillation is given by the equality
\begin{equation}
L/c = 2\pi \left( \frac{2p}{|m_{1}^{2}-m_{2}^{2}|} \right).
\label{14}
\end{equation}

The most important point in our analysis of the Gribov-Pontecorvo concept of neutrino oscillations is to find out the feasibility of the law of conservation of energy-momentum in the processes of production of electronic (anti-)neutrinos as a mixture (\ref{11}) of two (anti-)neutrinos $\nu_{1}$ and $\nu_{2}$ with different masses. By analogy with our assertion in Section 2.1, we compare the difference between the neutrino masses $\nu_{1}$ and $\nu_{2}$ with the total uncertainty in the masses of particles involved in the neutron decay process
\begin{equation}
n \rightarrow p + e^{-} + \bar{\nu}_{e}.
\label{15}
\end{equation}
In this process, the only unstable particle is the neutron with the lifetime value \cite{15} $\tau_{n} = 879$ s and, consequently, with the width $\Gamma_{n} = 7.5 \cdot 10^{-19}$ eV, which plays the role of the total uncertainty in masses. Energy-momentum conservation in the process (\ref{15}) is allowed if $|m_{1}-m_{2}| < \Gamma_{n}$. Let us take for the upper value of each of the masses $m_{1}$ and $m_{2}$ the experimentally established upper value of the electron neutrino mass \cite{15}: $m_{\nu_{e}} < 1.1$ eV. Then the possibility of conservation of energy-momentum in the process (\ref{14}) restricts the difference of squared neutrino masses $\nu_{1}$ and $\nu_{2}$ to the following limit
\begin{equation}
|m_{1}^{2}-m_{2}^{2}| < 8.2 \cdot 10^{-19} {\rm eV}^{2}. 
\label{16}
\end{equation}

Obviously, the restriction (\ref{16}) on the masses of the $\nu_{1}$ and $\nu_{2}$ components of the electron neutrino do not depend on the processes in which $\nu_{e}$ is produced. Let us now turn to the length $L$ of solar neutrino oscillations. For the observed solar neutrinos, as follows from the relation (\ref{14}, it is minimal at the threshold energy for gallium-to-germanium transitions equal to 0.233 MeV. Taking into account the inequality (\ref{16}), we have that $L > 7.0 \cdot 10^{14}$ km. Thus, the state of an electron neutrino on its way from the Sun to the Earth ($1.50\cdot 10^{8}$ km) could make no more than $2 \cdot 10^{-7}$ fractions of one oscillation. Such a situation would be absolutely unsuitable for solving the solar neutrinos problem.

{\it So, either the representation of the neutrino $\nu_{e}$ as a mixture (\ref{11}) has nothing to do with the solution of the solar neutrino problem, or in the case of the supremacy of the neutrino mass difference $|m_{1}-m_ {2}|$ over the neutron decay width ($7.5\cdot 10^{-19}$ eV) the energy-momentum conservation law would be violated, which makes such a representation and oscillations inadmissible.}

There have already been reasoning in the literature that the neutrino oscillation concept may contradict the law of conservation of energy-momentum, but neither the essence of the concept nor its effectiveness has been doubted. Thus, Giunti et al. \cite{18} believe that the question of energy–momentum nonconservation in processes involving neutrinos with an indeterminate mass does not arise in a real physical situation when the localization of the neutrino source and detector is taken into account. In fact, it is proposed to consider a neutrino on its way from the source to the detector as a virtual particle, the 4-momentum of which is outside the mass shell, which hides the certainty of the mass. Following this proposal, let's give the solar neutrino, registered on Earth, the status of a virtual one. Since virtuality lasts 500 s, due to the Heisenberg relation, the energy uncertainty is $1.3 \cdot 10^{-18}$ eV, which is very close to the discussed total mass uncertainty in the process (\ref{15}). Consequently, the conclusion formulated above remains fully valid, according to which either the survival of the electron component of the solar neutrino near the surface of the Earth differs very little from 1, or the electron neutrino cannot be a mixture of two other neutrinos with different masses and, therefore, its oscillations do not exist.

\begin{center}
{\bf 2.3. Wolfenstein's equation for neutrinos in a homogeneous medium}
\end{center}

Having established that the energy-momentum conservation law does not allow solving the solar neutrino problem on the basis of the neutrino oscillation hypothesis proposed by Gribov and Pontecorvo, we nevertheless continue to analyze its modifications.

The appearance of proposals to set up experiments on the detection of neutrinos from accelerators when they could pass through the Earth, along with ongoing experiments on the detection of solar neutrinos, prompted Wolfenstein to consider the possible influence of the medium on neutrino oscillations \cite{5}. When constructing his equation describing the transformation in the medium of probability amplitudes to detect neutrinos in the $\nu_{1}$ or $\nu_{2}$ state, Wolfenstein believed that coherent forward scattering of neutrinos on electrons takes place. This scattering is caused by electroweak interactions and gives different phase changes in the wave functions of the electron and muon components of the initial neutrino, which, in his opinion, is similar to the regeneration of $K_{S}$ from the $K_{L}$-beam passing through the medium.

I have two remarks with regard to Wolfenstein's assertion above.

First, we find reasoning about coherence in the work of Muller et al. \cite{19} devoted to the difference in the masses of neutral $K$-mesons. In it, coherence is associated with only one of the three types of regeneration of $K_{S}$ from $K_{L}$-mesons in a medium, when a plate placed in front of a parallel beam of $K_{L}$-particles generates a parallel beam of $K_{ S}$-particles. To two other types of regeneration, Muller et al. refer the diffraction scattering by nuclei and ordinary scattering by separate nucleons. The share of contributions to the total regeneration of each of its three types is not discussed in \cite{19}. From their experiment, Muller et al. found that the ratio $|m_{K_{L}} - m_{K_{S}}|/\Gamma_{K_{S}}$ is $0.85^{+0.3}_{-0.25 }$. From modern data \cite{15}, we obtain the value of $0.473$ for this ratio. Note, that subsequently Geveniger et al. \cite{20} measured the mass difference $m_{K_{L}} - m_{K_{S}}$ by the method of two regenerations and obtained a highly accurate value for it, essentially coinciding with the modern one. In \cite{20}, coherence is not mentioned.

Secondly, neutrino scattering on a certain set of electrons as a quantum mechanical phenomenon will be coherent in the case when the neutrino wave function has at the same time a non-zero value at the location of each electron from this set. This condition is satisfied if and only if the neutrino wavelength is comparable to the distance between electrons from a given set. Observable solar neutrinos with the lowest energy equal to 0.233 MeV have the longest wavelength $\lambda$, $\lambda = 8.47 \cdot 10^{-13}$ m. At the same time, electrons have the highest density $N_{e}$ in the center of the Sun, where the matter density, almost entirely determined by protons, is 148 g $\cdot$ cm$^{-3}$. There, the electron density $N_{e}$ is $8.85 \cdot 10^{25}$ cm$^{-3}$ and, consequently, the minimum distance between electrons in the Sun is $2.24 \cdot 10^{-11}$ m, which is approximately two orders of magnitude greater than the largest wavelength of the observed solar neutrino.

{\it So, coherence in the scattering of observed solar neutrinos is impossible. The analogy with the regeneration of $K$-mesons is weak and does not change this conclusion. Wolfenstein's equation is not applicable to any part of the trajectory of the observed solar neutrino. It must be attributed to some abstract medium and an abstract neutrino beam.}

\begin{center}
{\bf 2.4. Consequences of the Wolfenstein equation for solar neutrinos}
\end{center}

Let us put aside for a while the fact that the Gribov-Pontecorvo neutrino oscillation hypothesis and the Wolfenstein equation for neutrinos in a homogeneous medium are not suitable for describing the transformation of solar neutrinos, and analyze the consequences of extending the Wolfenstein equation to solar neutrinos.

The extension of the Wolfenstein equation to an inhomogeneous medium consists in giving the electron density the status of a function. In the $\nu_{e}, \;\nu_{\mu}$ basis, this equation has
the following form (see \cite{21}, Eq. (14.55))
\begin{eqnarray}
& &i\frac{d \;}{dt}\left(
\begin{array}{c}
A_{e}(t) \\
A_{\mu}(t)
\end{array} 
\right) = \frac{1}{2}\left(
\begin{array}{cc}
\displaystyle -\frac{\Delta m^{2}}{2E} \cos 2\theta + \sqrt{2} G_{F} N_{e}(t) & \displaystyle \frac{\Delta m^{2}}{2E} \sin 2\theta \\
\displaystyle \frac{\Delta m^{2}}{2E} \sin 2\theta & \displaystyle \frac{\Delta m^{2}}{2E} \cos 2\theta - \sqrt{2} G_{F} N_{e}(t)
\end{array}
\right) \times \nonumber \\
& & \times \left(
\begin{array}{c}
A_{e}(t) \\
A_{\mu}(t)
\end{array}
\right), \label{17}
\end{eqnarray}
where $A_{e}(t)$ ($A_{\mu}(t)$) is the probability amplitude that the neutrino state is electronic (muonic ) at the time moment $t$, when the electron density is equal to $N_{e}(t)$ at the location of the neutrino; $\Delta m^{2} = m_{1}^{2}- m_{2}^{2}$; $E$ is the neutrino energy.

It is easy to verify that the equation (\ref{17}) ensures the unitarity condition
\begin{equation}
|A_{e}(t)|^{2} + |A_{\mu}(t)|^{2} = 1.
\label{18}
\end{equation}

We first give a brief list of those analytical and numerical consequences of the equation (\ref{17}), which are detailed in our paper \cite{7}, adding some remarks.

We attribute the Wolfenstein equation to the entire time interval from the moment of the neutrino production in the Sun to its registration on the Earth, while the electron density at a neutrinos location can be quite large or arbitrarily small and zero value.

In our numerical analysis, we use the central values of the parameters of the two-neutrino model of solar neutrino oscillations shown by the Super-Kamiokande Collaboration \cite{22}:
\begin{equation}
\Delta m^{2} = 4.8^{+1.5}_{-0.8} \cdot 10^{-5} \; {\rm eV}^{2}, \qquad \sin^{2} \theta = 0.334^{+0.027}_{-0.023} .
\label{19}
\end{equation}
The conclusions in \cite{7} remain unchanged when the value of $\Delta m^{2}$ changes by two orders of magnitude, and the value of $\sin^{2} \theta$ changes twice.

Our solution to the equation (\ref{17}) is based on the adiabatic approximation used in quantum mechanics. Initially, it is assumed that, over several lengths of oscillations, the matter density in the Sun can be considered constant, equal to its value at the initial moment. After finding the interval of oscillation lengths corresponding to all observed values of the solar neutrino energy, we are convinced of the validity of this assumption when, using the numerical values of the dependence of the density of matter on its distance from the center of the Sun \cite{23}, we find that its relative change $\delta \rho / \rho$ throughout the length of one oscillation does not exceed $1.7 \cdot 10^{-3}$.

If at time moment $t_{0}$ the state of the solar neutrino was purely electronic, $|A_{e}(t_{0})|^{2} = 1$, then, as follows from the equation (\ref{17}), the probability $P_{e}(t)$ that this neutrino remains electron at time $t$ is given by
\begin{equation}
P_{e}(t) = |A_{e}(t)|^{2} = 1 - \sin^{2} 2\theta_{m} \sin^{2} f(N_{e})(t-t_{0})/2, 
\label{20}
\end{equation}
where
\begin{equation}
\sin 2\theta_{m} = \frac{\Delta m^{2}}{2E} \cdot \frac{\sin 2\theta}{f(N_{e})}, 
\label{21}
\end{equation}
\begin{equation}
f(N_{e}) = \sqrt{\left( \frac{\Delta m^{2}}{2E}\right) ^{2} -2\sqrt{2}\left( \frac{
\Delta m^{2}}{2E} \right) G_{F} N_{e} \cos 2\theta + 2 G_{F}^{2} N_{e}^{2}}.
\label{22}
\end{equation}
The angle $\theta_{m}$ defines the shape of the neutrinos $\nu_{e}$ and $\nu_{\mu}$ as a mixture of massive neutrinos in the medium $\nu_{1m}$ and $\nu_{2m}$:
\begin{equation}
\nu_{e} = \cos\theta\cdot\nu_{1m} + \sin\theta\cdot\nu_{2m}, \qquad
\nu_{\mu} = -\sin\theta\cdot\nu_{1m} + \cos\theta\cdot\nu_{2m}.
\label{23}
\end{equation}

The characteristics of each separate oscillation depend only on the density of the matter at the location of the neutrino and do not depend on the features of the part of the neutrino trajectory that preceded it. Knowledge of these characteristics allows us to get a conclusion about the set of all solar neutrino oscillations on any of their rectilinear trajectories for each value of the observed energy. At an energy in the range from 0.233 to 0.827 MeV, the length of the oscillations decreases monotonically with a decrease in the density of matter at the location of the neutrino. At an energy in the range from 0.827 to 18.8 MeV, the oscillation length as a function of the matter density has an extremum at its value
\begin{equation}
\rho_{\rm ext} = 1.95 \cdot 10^{-24} \frac{\cos 2\theta}{\sqrt{2} G_{F}} \left( \frac{\Delta m^{2}}{2E}\right) \; {\rm g}, 
\label{24}
\end{equation}
namely, the maximum
\begin{equation}
L(\rho)_{\rm max}/c = \frac{2\pi}{\sin 2 \theta} \left( \frac{2E}{\Delta m^{2}}\right),
\label{25}
\end{equation}
at that with increasing energy, the value of $\rho_{\rm ext}$ decreases from 148 to 6.51 g/cm$^ {3}$. The length of the oscillations of the observed solar neutrinos with energies from 0.233 to 18.8 MeV belongs to the interval from 12 to 1028 km, i.e. from $0.0000172 R_{0}$ to $0.00148 R_{0}$, where $R_{0}$ is the radius of the Sun, $R_{0} = 696000$ km.

When the matter density is $\rho_{\rm ext}$, the oscillating probability $P_{e}(t)$ has a maximum amplitude, namely, $P_{e}(t)$ can take any value from 1 to 0. Also, the maximum the value equal to 1 is taken by the quantity $\sin^{2} 2\theta_{m}$.

Based on a comparison of the lengths of neutrino oscillations in the Sun and in vacuum with the characteristic sizes of distributions of neutrino sources in the Sun, it was shown in \cite{7} that the numbers of oscillations along various trajectories from the Sun to the installation on Earth can differ by several tens or hundreds. As a result, the summation of the contributions of separate sources to the probability $P_{ee}$ of that the state of the solar neutrino will be electronic near the Earth's surface leads to the disappearance of durations of the neutrino motion along their trajectories, and the final formula for $P_{ee}$ for any neutrino energy has the following form
\begin{equation}
P_{ee} = 1-\frac{1}{2}\sin^{2} 2\theta. 
\label{26}
\end{equation}
The probability (\ref{26}) contradicts the results of at least three out of five experiments with solar neutrinos: on $^{37}{\rm Cl} \rightarrow ^{37}{\rm Ar}$ transitions, on elastic neutrino scattering on electrons, on the disintegration of deuterons by charged currents.

{\it So, three things are incompatible: the first is the values of the parameters of neutrino oscillations announced by the Super-Kamiokande and SNO collaborations, the second is the solution of the Wolfenstein equation for solar neutrinos with the indicated parameter values, the third is the experimentally measured rates of processes caused by solar neutrinos.}

\begin{center}
{\bf 2.5. Mikheev-Smirnov-Wolfenstein effect}
\end{center}

At the end of the 1970s, the problem of solar neutrinos acquired a clear numerical outline, consisting in the fact that the experimentally measured rate of the transition of chlorine into argon under the action of solar neutrinos is approximately 1/3 of the theoretical rate found in the framework of the standard solar model. For example, Davis announced \cite{24} that the measured rate of these transitions is $2.2 \pm 0.3$ SNU, and Bahcall with colleagues reported \cite{25} that the calculated rate is $7.5 \pm 1.5$. At the same time, for any values of the parameters and two- and three-neutrino oscillations in their standard sense, the probability of that the initial electron neutrino remains electronic at the place of its registration, when averaged over the time of one oscillation, is no less than 1/2.

Under these conditions, an extraordinary scenario of neutrino oscillations was announced by Mikheev and Smirnov in \cite{6}, the main assertion of which is that the initial beam of electron neutrinos, after passing through a certain layer of matter, is almost completely transformed into a beam of muon neutrinos. The substantiation of this assertion is not accompanied by any analytical calculations and formulas. Its role is played by the terminology that is not appropriate in particle physics and field theory, the haziness of statements, the purposeful game with the dependence of the mixing angle on the density of the medium, which plays a key role in declaring the metamorphosis of the solar neutrino.

Let us dwell on all the details of the work \cite{6}, taking into account that it has hundreds of adherents.

Instead of speaking, as it is customary in mathematical analysis, about the existence of extremal values of a number of oscillation characteristics, as functions of the density of matter, on a certain set of neutrino trajectories in the Sun, Mikheev and Smirnov declare the existence of a resonant amplification of neutrino oscillations in the medium. The term "resonance" in the oscillatory process has both scientific and everyday use. In mechanics and in the theory of electricity, the appearance of resonance in oscillations is associated with the existence of three constituent elements: the frequency of proper oscillations, the frequency of external influences and "the friction" (energy loss). Only the first of these elements is present during neutrino oscillations in the medium. The everyday understanding of resonance is associated with the possible dramatic consequences of vibrations (for example, with the destruction of a bridge).

Since "resonance" is associated with the quantity $\sin^{2} 2\theta_{m}$ and with the amplitude of oscillations as functions of the density of matter, then its "width" is considered. These functions have values greater than half of their maximum values, equal to 1, in the following interval of matter density
\begin{equation}
\Delta \rho_{\rm ext} \in [{\rm max}(\rho_{\rm ext}(1 - \tan \theta),0) \; \rho_{\rm ext}(1 + \tan \theta)].
\label{27}
\end{equation} 
The layer of solar matter, the density in which belongs to the interval (\ref{27}), is called the resonance layer by Mikheev and Smirnov. Let's estimate its size. At a solar neutrino energy of 6 MeV, the extreme density of matter $\rho_{\rm ext}$ is, according to the relation (\ref{24}), 20.4 g$\cdot$cm$^{-3}$ and is at a distance of 0.254$ R_{0}$ from the center of the Sun. The density in the resonant layer (\ref{27}) ranges from 0 to 84.3 g$\cdot$cm$^{-3}$ and is at a distance from $R_{0}$ to 0.102$R_{0}$ from the center Sun. Hundreds or thousands of solar neutrino oscillations can occur in this resonant layer. In addition to the term "resonance layer", the term "final object" with a matter density from 0 to $\rho_{\rm max}$ (without specification) or simply "object" is also used. The selection of separate layers of the Sun does not have any logical consequences, but rather looks like some kind of ritual.

Before analyzing the assertion about the transformation of an electron neutrino into a muon neutrino when passing through a medium, let us fix an unambiguous interpretation of such a transformation. Since the probability of detecting a neutrino being in the $\nu_{e}$ state at a certain moment of time can be either greater or less than the probability of detecting a neutrino being in the $\nu_{\mu}$ state at the same time moment, the change in the ratio of these probabilities does not can be a criterion for transformation. There are two stable indications that the oscillating state of the neutrino must be considered either electronic or muon: one is the superiority of the average probability of detecting one type of neutrino over the average probability of detecting another type over the time of one oscillation, and the other is the superiority of the amplitude of the probability oscillations for one type over the amplitude for another type. During the entire time allotted to the Wolfenstein equation for solar neutrinos, the above relations remain unchanged, i.e. transformation of one type of neutrino into another is impossible.

Let us now pay attention to the character of the changes in the value of $\sin 2\theta_{m}$, given by the relations (\ref{21}) and (\ref{22}), and to two possible variants of changes in the mixing angle $\theta_{m}$ for neutrinos with a certain energy along one or another trajectory. At neutrino energies from 0.233 to 0.827 MeV, when there is no extremal point on the trajectory, the value of $\sin\theta_{m}$ increases with increasing matter density, starting from $\sin 2\theta$ and not reaching 1, and the angle $2\theta_{m}$ grows starting from $2\theta$ and not reaching $\pi/2$. At energies above 0.827 MeV, the value of $\sin 2\theta_{m}$ first increases from $\sin 2\theta$ to 1, and then decreases to a certain value $c_{0}$, which is the smaller, the higher the matter density at the production place of the neutrino. The first variant of changing the angle $2\theta_{m}$ is following: as the matter density increases, it grows from $2\theta$ to $\pi/2$, and then decreases to the value $\arcsin c_{0}$. The second variant is next: the angle $2\theta_{m}$ with increasing matter density invariably increases from $2\theta$ to $\pi - \arcsin c_{0}$.

In \cite{6}, a variant of changing the angle $\theta_{m}$ from $\theta$ to $\pi/2 - (\arcsin c_{0})/2$ is chosen without mentioning the existence of the first variant. In the case of small values of $\theta$ and $c_{0}$, the chosen variant means a change in the mixing angle along the neutrino trajectory by approximately $\pi/2$, while in the first variant such a change is small. Now comes the climax of the metamorphose, described fragmentarily and muzzily both by Mikheev and Smirnov in \cite{6} and by Smirnov in a recent report \cite{26}, which I present in an orderly and complete form. For a small value of $c_{0}$, the first of the relations (\ref{23}) at the moment of production of the solar neutrino, when it is purely electronic, is written as
$$\nu_{e} = \cos [\pi/2-(\arcsin c_{0})/2] \cdot \nu_{1m}(\rho_{\rm max}) + 
\sin [\pi/2-(\arcsin c_{0})/2] \cdot \nu_{2m}(\rho_{\rm max})$$ 
\begin{equation}
\approx [c_{0}/2]\cdot \nu_{1m}(\rho_{\rm max})+
\nu_{2m}(\rho_{\rm max})
\label{29}
\end{equation}
The transformation of the state of a neutrino during its motion in the Sun is completely determined by the changes in the mixing angle $\theta_{m}$ and the states of massive neutrinos in matter on the right side of the first of the relations (\ref{23}). At the moment when the solar neutrino is near the exit from the Sun, its state $\nu_{\rm sol}$ is given by the formula
\begin{equation}
\nu_{sol} = \cos \theta \cdot \nu_{1m}(\rho=0) + \sin \theta \cdot \nu_{2m}(\rho=0).
\label{30}
\end{equation}
Since the expansion (\ref{29}) is completely dominated by the neutrino $\nu_{2}$, and the expansion (\ref{30}) is dominated by the neutrino $\nu_{1}$, Mikheev and Smirnov conclude that the final neutrino is muon, and the probability amplitude to find $\nu{e}$ in the final state is equal to $\sin \theta$.

The significant difference in the values of the coefficients related to the same massive neutrinos in the decompositions (\ref{29}) and (\ref{30}), enhanced in \cite{6} by the smallness of the angle $theta$ accepted there, is entirely due to the meaningful by choosing the second variant of the dependence of the mixing angle $\theta_{m}$ on the density of the medium. 

In the first variant of the dependence of the mixing angle $\theta_{m}$ on the density of the medium, instead of the relation (\ref{29}), we have the following equality
\begin{equation}
\nu_{e} = \cos [(\arcsin c_{0})/2] \cdot \nu_{1m}(\rho_{\rm max}) +
\sin [(\arcsin c_{0})/2] \cdot \nu_{2m}(\rho_{\rm max}) \approx \nu_{1m}(\rho_{\rm max})+
[c_{0}/2] \cdot \nu_{2m}(\rho_{\rm max})
\label{29a}
\end{equation}

Now, in both expansions (\ref{29a}) and (\ref{30}), the same neutrino $\nu_{1}$ dominates and, following Mikheev and Smirnov, one should say that both the initial and final neutrinos are electronic. The conclusion that the final neutrino is a muon neutrino vanishes like a dream.

The assertion of Mikheev and Smirnov about the transformation of an electron neutrino into a muon neutrino when passing through a medium turns out to be primitive false, moreover, the notions of "resonance", "resonance layer", "adiabatic regime" turned out to be unnecessary for its imaginary substantiation.

For many years I was surprised that the paper by Mikheev and Smirnov \cite{6} discussed here was not rejected in the journal Yadernaya Fizika. I unexpectedly received an answer from Smirnov, who outlined some points from the conceiving of the Mikheev-Smirnov-Wolfenstein mechanism in a report at a conference in Paris in 2018 \cite{26}. It turns out that paper \cite{6} was initially rejected in the journal Yadernaya Fizika. Smirnov writes that "it was a general skepticism: 'something should be wrong, somebody will eventually find this'". Pontecorvo  also had a skeptical opinion (it is "rubbish") (section 3.6 of \cite{26}). However, thanks to the influence of Academician G.T. Zatsepin, the article was accepted by Yadernaya Fizika and also published in Nuovo Cimento. Later, Pontecorvo together with Wolfenstein "concluded that, it seems, there is no practical outcome of oscillations in matter" (section 3.5 of \cite{26}).

{\it So, the key thesis of the Mikheev-Smirnov-Wolfenstein mechanism about the transformation of an electron neutrino into a muon neutrino when passing through the solar medium is erroneous, which makes this mechanism generally unsuitable for solving the solar neutrino problem.}

\begin{center}
{\large \bf 3. New fundamental interaction}
\end{center}
\begin{center}
{\bf 3.1. Solving the solar neutrino problem}
\end{center}

The search by many diverse specialists for the reasons for the discrepancy between the result of the Davis experiment \cite{1} and Bahcall's prediction \cite{2}, reported by Reines \cite{3}, was not successful. It, of course, was conducted within the framework of the standard model of interactions. This alone serves as a sufficient reason to go beyond the standard model and, adhering to classical principles in physics, postulate the existence of a new interaction that is not a consequence of known interactions, and therefore is fundamental. The essence of the new interaction will not suffer, if we do not consider it fundamental.

In \cite{8}, we put forward and substantiated the hypothesis that the emergence of the solar neutrino problem is due to the existence of a rather hidden interaction, which is carried by a massless pseudoscalar boson having Yukawa coupling with an electron neutrino, proton, and neutron (with u- and d-quarks), reflected by the following relativistically invariant Lagrangian
\begin{equation}
 L = ig_{\nu_{e}ps}\bar{\nu}_{e}\gamma^{5}\nu_{e}\varphi_{ps}+
ig_{Nps}\bar{p}\gamma^{5}p\varphi_{ps}-ig_{Nps}\bar{n}\gamma^{5}n\varphi_{ps},
\label{31}
\end{equation}
and not coupled with the electron at the tree level. We do not rule out that the coupling of the pseudoscalar boson with different types of neutrinos may be different, if it exists at all.

The decrease in the experimental rate of observed processes with solar neutrinos compared to the theoretical one is caused by two reasons. The first reason is the change in its handedness from left to right and vice versa at each collision of a neutrino with a nucleon, which is due to the Lorentzian structure of the pseudoscalar current:
\begin{equation}
\bar{\psi} (p_{2}) \gamma^{5} \psi (p_{1}) = \bar{\psi}_{R} (p_{2}) \gamma^{5} \psi_{L} (p_{1}) - \bar{\psi}_{L} (p_{2}) \gamma^{5}_{R} \psi (p_{1}).
\label{32}
\end{equation} 
The second reason is the decrease in the neutrino energy $\omega$ at collision with a resting nucleon of mass $M$ by the value $\Delta \omega$ uniformly distributed in the following interval
\begin{equation}
\Delta \omega \in [0, \frac{2\omega^{2}}{M}\cdot \frac{1}{1+2\omega/M}].
\label{33}
\end{equation} 

The flux of solar neutrinos near the surface of the Earth does not differ from the theoretical one. But due to the first reason, the flux of left-handed neutrinos decreases by about a factor of two compared to the theoretical one: its ratio to the flux of right-handed neutrinos is, as it shown in \cite{10}, 0.516:0.484 (after correction here, it is 0.515:0.485).

As is well known, the cross sections for the interaction of a neutrino with electrons, nucleons, and nuclei increase significantly with increasing energy up to its value for gauge bosons, but increase differently near the thresholds of various processes. Therefore, a decrease in the energy of each solar neutrino entails a decrease in the rate of observed processes, which is different for different processes.

The solution of the task about the consequences of the short-term Brownian motion of neutrinos in the Sun was facilitated by the fact that the cross section of elastic scattering on a nucleon at rest, caused by the interaction (\ref{31}),
\begin{equation}
\sigma = \frac{(g_{\nu_{e}ps}g_{Nps})^{2}}{16\pi M^{2}} \frac{1}{1+2\omega/M}]
\label{34}
\end{equation}
is practically independent of the energy of the solar neutrino. Therefore, with sufficient accuracy, we can assume that the distributions of the number of collisions of neutrinos with solar nucleons are the same for all solar neutrinos.

In \cite{8} and \cite{9}, the role of a consequence of the unknown distribution is played by the effective number $n_{0}$ of collisions of neutrinos with nucleons of the Sun, which determines the energy spectrum of neutrinos near the Earth's surface, regardless of their handedness. Focusing on numerical methods dealing with a discrete set of numbers, we consider that the energy of the scattered neutrino takes with equal probability the two extreme values of its allowable interval, i.e. that the initial energy level of a neutrino after its collision with a nucleon splits into two levels. After $n_{0}$ collisions of neutrinos with nucleons, its initial energy level $\omega$ splits into $n_{0}+1$ binomially distributed levels:
\begin{equation}
\omega_{1} = \omega, \quad \omega_{2} = \frac{\omega_{1}}{1+2\omega_{1}/M}, \quad \ldots
\omega_{n_{0}+1} = \frac{\omega_{n_{0}}}{1+2\omega_{n_{0}}/M}.
\label{35}
\end{equation}

In \cite{9}, simple algorithms for calculating the numerical values of the rates of each of the five observed processes with solar neutrinos are given, as well as references to the works of Bahcall et al., which contain tabulated values of solar neutrino fluxes from their various sources and cross sections for a number of processes. This makes it easy for any physicist or mathematician to check our results. Now I recommend replacing the multiplier $0.5 \Phi$ by $0.515 \Phi$ in the formulas of \cite{9} for the rates of processes. Good agreement between the theoretical results and the results of all five experiments with solar neutrinos is ensured at $n_{0} = 12$.

The small number of collisions of neutrinos with nucleons of the Sun allows us to consider it plausible that one neutrino collision occurs when a neutrino passes by about 0.7 to 0.9 fractions of the maximum possible number of nucleons on its way in the Sun. Taking into account the distribution of matter in the Sun given in \cite{23} and the relation (\ref{34}), we have obtained the following estimate
\begin{equation}
\frac{g_{\nu_{e}ps}g_{Nps}}{4\pi} = (3.2 \pm 0.2) \cdot 10^{-5}.
\label{36}
\end{equation}
This estimate is used in calculating the deuteron disintegration rate by neutral currents caused by both left- and right-handed  solar neutrinos, $\nu_{e}+D \rightarrow \nu_{e}+n+p$.

Each of the constants $g_{\nu_{e}ps}$ and $g_{Nps}$ individually can a priori have values in a fairly large range. Meanwhile, processes in which the massless pseudoscalar boson $\varphi_{ps}$ interacts only with nucleons or only with electron neutrinos present undoubted interest both for the physics of the Sun and for the physics of the Earth.

Later, in \cite{10}, the Brownian motion of neutrinos in the Sun was successfully described by the geometric distribution of the number of their collisions with nucleons
\begin{equation}
P(n) = p(1-p)^{n}, \qquad n = 0,1,2,\ldots. 
\label{37}
\end{equation}
From a comparison of the theoretical results based on this distribution with the experimental results in \cite{10}, it was accepted that the parameter $p$ is equal to 0.062. Subsequently, I again made such a comparison and now I cosider that $p=0.060$. The theoretical rates of all five observed processes given below were calculated exactly at $p=0.060$.

In accordance with the fact that the four observed processes, $\nu_{e}+^{37}{\rm Cl} \rightarrow e^{-}+^{37}{\rm Ar}$, $\nu_{e }+^{71}{\rm Ga} \rightarrow e^{-}+^{71}{\rm Ge}$, $\nu_{e}+e^{-} \rightarrow nu_{e}+e ^{-}$ and $\nu_{e}+D \rightarrow e^{-}+p+p$, and subprocess $\nu_{e}+D \rightarrow \nu_{e}+n+p$ with $Z$-boson exchange, are caused only by left-handed neutrinos, then when calculating the theoretical rates of these processes, the contributions of only those solar neutrinos that have undergone an even number of collisions with solar nucleons are summed up. The subprocess $\nu_{e}+D \rightarrow \nu_{e}+n+p$ with the exchange of pseudoscalar boson $\varphi_{ps}$, is contributed by both left- and right-handed neutrinos that have experienced any number of collisions with nucleons of the Sun.

   Tables 1-5 below show the results of calculating the velocities of all observed processes with solar neutrinos based on the method of the effective number of neutrino-nucleon collisions $n_{0}$ for two values of the probability $W_{\rm left} $ that at the Earth's surface the solar neutrino is left-handed, and based on geometric distribution (\ref {4}). Tables 3-4 show the dependence of the rates of the processes on the lower value $E_{c}$ of the reconstructed energy of the final electron.

\begin{center}
\begin{tabular}{lccccccc}
\multicolumn{8}{c}{{\bf Table 1.} The rate of transitions ${}^{37}{\rm Cl} \rightarrow {}^{37}{\rm Ar}$ in SNU.} \\
\hline
\multicolumn{1}{l}{} 
&\multicolumn{1}{c}{${}^{8}{\rm B}$} 
&\multicolumn{1}{c}{${}^{7}{\rm Be}$}
&\multicolumn{1}{c}{${}^{15}{\rm O}$}
&\multicolumn{1}{c}{$pep$}
&\multicolumn{1}{c}{${}^{13}{\rm N}$}
&\multicolumn{1}{c}{$hep$}
&\multicolumn{1}{c}{Total} \\ 
\hline
Experiment \cite{27} &  &  &  &  &  &  & $2.56 \pm 0.16 \pm 0.16$ \\
$W_{\rm left}=0.5$, $n_{0} = 11$ & 1.97 & 0.43 & 0.17 & 0.11 & 0.04 & 0.01 & 2.72 \\
$W_{\rm left}=0.515$, $n_{0} = 12$ & 1.95 & 0.43 & 0.17 & 0.11 & 0.05 & 0.01 & 2.72 \\
Eq. (\ref{37}), $p=0.060$ & 1.99 & 0.41 & 0.17 & 0.11 & 0.04 & 0.01 & 2.73 \\
\hline
\end{tabular}
\end{center}

\begin{center}
\begin{tabular}{lcccccccc}
\multicolumn{9}{c}{{\bf Table 2.} The rate of transitions ${}^{71}{\rm Ga} \rightarrow {}^{71}{\rm Ge}$ in SNU.} \\ 
\hline
\multicolumn{1}{l}{}
&\multicolumn{1}{c}{$p$-$p$}
&\multicolumn{1}{c}{${}^{7}{\rm Be}$} 
&\multicolumn{1}{c}{${}^{8}{\rm B}$} 
&\multicolumn{1}{c}{${}^{15}{\rm O}$}
&\multicolumn{1}{c}{${}^{13}{\rm N}$}
&\multicolumn{1}{c}{$pep$}
&\multicolumn{1}{c}{$hep$}
&\multicolumn{1}{c}{Total} \\ 
\hline
Experiment \cite{28} &  &  &  &  &  &  &  & $62.9^{+6.0}_{-5.9}$ \\
Experiment \cite{29} &  &  &  &  &  &  &  & $65.4^{+3.1}_{-3.0}{}^{+2.6}_{-2.8}$ \\
$W_{\rm left}=0.5$, $n_{0} = 11$ & 34.6 & 17.2 & 4.9 & 2.8 & 1.7 & 1.4 & 0.02 & 62.6 \\
$W_{\rm left}=0.515$, $n_{0} = 12$ & 35.7 & 17.7 & 4.9 & 2.9 & 1.7 & 1.4 & 0.02 & 64.4 \\
Eq. (\ref{37}), $p=0.060$ & 35.5 & 17.5 & 4.9 & 2.8 & 1.7 & 1.4 & 0.02 & 63.8 \\
\hline
\end{tabular}
\end{center}

\begin{center}
{{\bf Table 3.} Effective fluxes of neutrinos $\Phi_{eff}^{\nu e}({}^{8}{\rm B})$ found from the process} \\
\begin{tabular}{lccccc}
\multicolumn{6}{l}{$\nu_{e} e^{-}\rightarrow \nu_{e} e^{-}$ ($E_{c}$ is given in MeV, and the fluxes are in units of $10^{6}$ ${\rm cm}^{-2}{\rm s}^{-1}$).} \\  
\hline
\multicolumn{1}{l}{References}
&\multicolumn{1}{c}{$E_{c}$} 
&\multicolumn{1}{c}{Experimental}
&\multicolumn{1}{c}{$W_{\rm left}=0.5,$} 
&\multicolumn{1}{c}{$W_{\rm left}=0.515,$}
&\multicolumn{1}{c}{Eq. (\ref{37}),} \\
&& results & {$n_{0} = 11$} & {$n_{0} = 12$} & {$p=0.060$} \\
\hline
SK IV \cite{22} & 4.0 & $2.31\pm 0.02 \pm 0.04$ & 2.39 & 2.42 & 2.37 \\
SNO I \cite{30} & 5.5 &$2.39^{+0.24}_{-0.23}{}^{+0.12}_{-0.12}$ & 2.19 & 2.19 & 2.15\\
SNO II \cite{31} & 6.0 &$2.35\pm 0.22\pm 0.15$ & 2.10 & 2.10 & 2.07\\
SNO III \cite{32} & 6.5 &$1.77^{+0.24}_{-0.21}{}^{+0.09}_{-0.10}$ & 2.01 & 2.00 & 1.98\\
\hline
\end{tabular}
\end{center}

\begin{center}
{{\bf Table 4.} Effective fluxes of neutrinos $\Phi_{eff}^{cc}({}^{8}{\rm B})$ found from the process} \\
\begin{tabular}{lccccc}
\multicolumn{6}{l}{$\nu_{e}D \rightarrow  e^{-}pp$ ($E_{c}$ is given in MeV,and the fluxes are in units of $10^{6}$ ${\rm cm}^{-2}{\rm s}^{-1}$).} \\ 
\hline
\multicolumn{1}{l}{References}
&\multicolumn{1}{c}{$E_{c}$} 
&\multicolumn{1}{c}{Experimental}
&\multicolumn{1}{c}{$W_{\rm left}=0.5,$} 
&\multicolumn{1}{c}{$W_{\rm left}=0.515,$}
&\multicolumn{1}{c}{Eq. (\ref{37}),} \\
&& results & {$n_{0} = 11$} & {$n_{0} = 12$} & {$p=0.060$} \\
\hline
SNO I \cite{30} & 5.5 & $1.76^{+0.06}_{-0.05}{}^{+0.09}_{-0.09}$ & 1.86 & 1.84 & 1.85 \\
SNO II \cite{31} & 6.0 & $1.68^{+0.06}_{-0.06}{}^{+0.08}_{-0.09}$ & 1.77 & 1.73 & 1.77 \\
SNO III \cite{32} & 6.5 & $1.67^{+0.05}_{-0.04}{}^{+0.07}_{-0.08}$ & 1.66 & 1.62 & 1.69 \\
\hline
\end{tabular}
\end{center}

\begin{center}
{{\bf Table 5.} Effective fluxes of neutrinos $\Phi_{eff}^{nc}({}^{8}{\rm B})$ found from the process} \\
\begin{tabular}{lccccc}
\multicolumn{6}{c}{$\nu_{e}D \rightarrow  \nu_{e}np$ (the fluxes are in units of $10^{6}$ ${\rm cm}^{-2}{\rm s}^{-1}$).} \\
\hline
\multicolumn{1}{l}{}
&\multicolumn{1}{c}{Exchange}
&\multicolumn{1}{c}{Exchange}
&\multicolumn{1}{c}{Exchange} 
&\multicolumn{1}{c}{Exchange}
&\multicolumn{1}{c}{Sum} \\
& by $Z$ & by $Z$ & by $Z$ & by $\varphi$ & \\
& $W_{\rm left}=0.5$ & $W_{\rm left}=0.515$ & Eq. (\ref{37}) & all neutrinos & \\
\hline
SNO I \cite{30} & & & & & $5.09^{+0.44}_{-0.43}{}^{+0.46}_{-0.43}$ \\
SNO II \cite{31} & & & & & $4.94^{+0.21}_{-0.21}{}^{+0.38}_{-0.34}$  \\
SNO III \cite{32} & & & & & $5.54^{+0.33}_{-0.31}{}^{+0.36}_{-0.34}$ \\
\hline
$n_{0} = 11$ & 2.10 & & & 2.87 & 4.98 \\
$n_{0} = 12$ & & 2.11 & & 2.85 & 4.96 \\
Eq. (\ref{37}) & & & 2.10 & 2.79 & 4.89 \\
\hline
\end{tabular}
\end{center}

Comparing the corresponding numbers in each of the tables, we come to the following conclusion. The results of collisions of neutrinos with nucleons in the Sun caused by the interaction (\ref{31}), which were obtained on the basis of the method of the effective number of collisions with a slight superiority of the flux of left-handed neutrinos, and which were obtained based on the geometric distribution of collisions, are close to each other and agree well with the experimental results. 

{\it A good agreement of theoretical results, obtained on the basis of the hypothesis of the existence of a new fundamental interaction, with the results of five absolutely different experiments with solar neutrinos can only be a game of chance with negligible probability.}

\begin{center}
{\bf 3.2. Borexino, KamLAND}
\end{center}

There are two types of Borexino experiments. First, the elastic scattering of solar neutrinos on electrons at the scattered electron energy greater than 3 MeV is studied. The observed effective flux of solar neutrinos from $^{8}$B \cite{33} is consistent with the experimental results of Super-Kamiokande and SNO. As this experiment was carried out after the corresponding experiments of Super-Kamiokande and SNO, and as the accuracy of its result is much less than in Super-Kamiokande and SNO, then it adds nothing new to the solar neutrinos. Second, the solar neutrino flux from $^{7}$Be is studied. However, as it is stated in the work \cite{34}, "the contribution of the pp, pep, CNO, and $^{8}$B solar neutrino were fixed to the SSM-predicted rates assuming MSM neutrino oscillations." Thus, Borexino does not present a purely 
experimental result for the neutrino flux from $^{7}$Be, but only an MSM interpretation of it, and so, we have nowhere to apply our theoretical calculations.

The preparation and implementation of the KamLAND experiment was initially focused (see Ref. \cite{35}) on confirming the existence of electron (anti-)neutrino oscillations. When presenting the first results, it was said \cite{36}: "The long baseline, typically 180 km, enables KamLAND to address the oscillation solution of the ‘solar neutrino problem’ using reactor anti-neutrinos under laboratory conditions". The existence of (anti-)neutrino oscillations with parameters given by the KamLAND collaboration and close to ones of the SNO and Super-Kamiokande collaborations would lead
to the survival probability of the solar neutrino electron component near the Earth's surface, given by the formula (\ref{26}), which, as stated in section 2.4, contradicts at least three out of five experiments. At the same time, the interaction (\ref{31}), which provided a solution to the solar neutrino problem, has nothing to do with the announced result of the KamLAND collaboration, since the reactor antineutrino cannot experience a single collision on its 180 km path.

After obtaining a solution to the solar neutrino problem, we had no doubt that the reason for the loss of observed events in the KamLAND experiment is serious omissions in its setting up and processing. I have repeatedly returned to the issue of such omissions, having studied, where possible, the features of the appearance of fluorescent light, the characteristics of its spectrum, the dependence on the wavelength of the fluorescent signal of the attenuation of its intensity during propagation in liquid KamLAND scintillator, etc. Based on the comparison of suitable numbers, I finally came to the following conclusion \cite{37}. In the conditions of unprecedentedly large sizes of KamLAND detector and of the refusal of using the gadolinium, the absence of (1) proper study of the real spectrum of the PPO fluorescent light, of (2) a detailed study of the fluorescent light attenuation in a liquid scintillator depending on its wavelengths, of (3) the influence on the attenuation of light of its scattering and re-emission, and of (4) the theoretical calculation of the observability of events $\bar{\nu}_{e} + p \rightarrow  e^{+} + n$ opens the door to uncontrolled loss of such events. Therefore, we do not consider the announced results of the KamLAND experiment as reliable.

\begin{center}
{\bf 3.3. Reactor antineutrino anomaly}
\end{center}

Acquisition of knowledge about the spectrum of reactor antineutrinos is realized by two different methods. One of them, called experimental, uses the spectrum of observed positrons produced in an experimental setup under the action of reactor antineutrinos, and the well-known dependence of the inverse beta decay cross section on antineutrino energy. Another method, called theoretical, is based on finding electron spectra in the beta decays of $^{235}$U, $^{239}$Pu, $^{241}$Pu and $^{238}$U and then on converting them into antineutrino spectra. This began with experiments at the Grenoble reactor \cite{38} and with the work of Schreckenbach \cite{39}. Subsequently, theoretical calculations of the spectra of electrons and antineutrinos in thousands of beta decay branches of the above-mentioned nuclei were performed \cite{40}. Knowledge of the summary antineutrino spectra in the beta decays of each of these nuclei, obtained as a result of converting the electron spectra, allowed to find the expected rate of inverse beta decay events in a given reactor and compare it with the results of similar events in the entire set of experiments with distances less than 100 meters from the reactors, the authors of \cite{41} conclude that the ratio of the rate of observed events to the rate of predicted events is equal to $0.943 \pm 0.023$. The distinction of this ratio from unity is called by them the reactor antineutrino anomaly. The authors of \cite{41} are inclined to believe that the reactor anomaly may indicate the existence of a fourth nonstandard neutrino, which determines electron neutrino oscillations at short distances.

The existence of interaction (\ref{31}) makes difference between the theoretical and experimental spectra of antineutrinos produced in beta decays of any nuclei inevitable. The magnitude of this difference is determined by the Yukawa coupling constant of a massless pseudoscalar boson with an electron neutrino $g_{\nu_{e}ps}$, which until now remained unknown.

Once again, we note that the experimental spectrum includes those and only those antineutrinos that manifest themselves in the production of positrons. The theoretical spectrum, relied on the spectrum of electrons, includes both those antineutrinos that are capable of producing a positron, and those that do not have such an ability. The interaction (\ref{31}) allows the existence along with the dominant mode of beta decay of the X nucleus into the Y nucleus
\begin{equation}
{\rm X} \rightarrow {\rm Y} + e^{-} + \bar{\nu}_{eR},
\label{39} 
\end{equation}
also of the decay mode with the emission of a massless pseudoscalar boson
\begin{equation}
{\rm X} \rightarrow {\rm Y} + e^{-} + \bar{\nu}_{eL} + \varphi_{ps}. 
\label{40}
\end{equation}
The right-handed antineutrino, born in the beta decay of the X nucleus, can emit a massless pseudoscalar boson in a virtual state, changing its handedness to the left. The left-handed antineutrino within the framework of the standard interaction cannot manifest itself, causing the production of a positron. Thus, the experimental spectrum includes only antineutrinos of the dominant nuclear beta-decay mode (\ref{39}). The theoretical spectrum includes both antineutrinos from the dominant mode (\ref{39}) and effective "antineutrino" that accompanies the electron and is the sum of the unobservable left-handed antineutrino and the unobservable massless pseudoscalar boson from the mode (\ref{40}).

To estimate the order of magnitude of the constant $g_{\nu_{e}ps}$, we calculated the ratios of the probability of beta decay mode (\ref{40}) to the probability of main mode (\ref{39}) for the two beta decay branches, extending from the nucleus $^{235}$U, noted in \cite{42}, namely: beta transition of $^{143}$Xe to $^{143}$Cs and beta transition of $^{90}$ Br to $^{90}$Kr. The details of the calculations will be presented elsewhere, but for now we present only some resumes.

To eliminate the massless divergence in the probability of beta decay mode (\ref{40}), the quantity $\lambda$ is introduced. The emission of a massless pseudoscalar boson provides the level of ratio of the observed and expected rates of inverse beta decay in reactors, $0.943 \pm 0.023$, announced in \cite{41}, if at the value $\lambda = 10^{-3}$ eV the following estimate takes place 
\begin{equation}
g_{\nu_{e}ps} = 0.12 \pm 0.05, 
\label{41}
\end{equation}
 which is the same for both beta decay branches considered. The estimate of the constant $g_{\nu_{e}ps}$ depends weakly on the value of the parameter $\lambda$. Thus, at $\lambda = 10^{-6}$ eV we obtain: $g_{\nu_{e}ps} = 0.104 \pm 0.04$ .

At the same time, we obtain an estimate for the Yukawa coupling constant of a massless pseudoscalar boson with nucleons. Namely, taking into account equality (\ref{36}), we have at the value $\lambda = 10^{-3}$ eV
\begin{equation}
g_{Nps} = (8.5 \pm 3.8) \cdot 10^{-9}. 
\label{42}
\end{equation}

\begin{center}
{\large \bf 3.4. Around the gallium anomaly}   
\end{center}

The notion of gallium anomaly, which appeared earlier than the notion of reactron antineutrino anomaly, means the difference between the numbers of observed and expected events of transitions of $^{71}$Ga to $^{71}$Ge under the influence of electron neutrinos from the decay of $^{51}$Cr. At present, there is a scatter in the ratio of these numbers obtained in six variants of completed experiments, in the range from $0.95 pm 0.12$ to $0.77 \pm 0.05$ \cite{43}, \cite{44}. 

The appearance of the named difference associated with the decay of $^{51}$Cr is inevitable due to the emission of a massless pseudoscalar boson from neutrinos, as a result of which, along with the main electron capture mode
\begin{equation}
^{51}{\rm Cr} + e^{-} \rightarrow ^{51}{\rm V} + \nu_{eL}
\label{43}
\end{equation}
there is another mode
\begin{equation}
^{51}{\rm Cr} + e^{-} \rightarrow ^{51}{\rm V} + \nu_{eR} +\varphi_{ps}.
\label{44}
\end{equation} 
that cannot lead to transitions of $^{71}$Ga to $^{71}$Ge.

In order to find the level of the gallium anomaly, taking into account the equality (\ref{41}), it is necessary to carry out accurate calculations of the cross sections of both indicated electron capture modes. Apparently, it would not be amiss to re-analyze all the theoretical and experimental aspects of setting up an experiment with $^{51}$Cr.

\begin{center}
{\large \bf 4. Conclusion}
\end{center}

Ignoring the basic principles of classical logic, researchers lose the ability to distinguish between perfect reasoning and charlatanis, what is inherent in the concept of particle oscillations from its inception to the present day.

The solution to the problem of solar neutrinos, and then clarifying the nature of reactor antineutrino anomaliy, is provided  by classical field theory methods with their inherent simplicity.

\end{small}
\end{document}